\documentclass[12pt]{iopart}

\usepackage{iopams}
\usepackage{graphicx}
\begin{document}

\title{Analytical expression for a class of spherically symmetric solutions in Lorentz breaking massive gravity}

\author{Ping Li, Xin-zhou Li and Ping Xi}

\address{Center for Astrophysics,Shanghai Normal University,100 Guilin Road,Shanghai,200234,China}
\ead{\mailto{li57120@126.com}, \mailto{kychz@shnu.edu.cn} and \mailto{xiping@shnu.edu.cn}}
\vspace{10pt}
\begin{indented}
\item[]
\end{indented}

\begin{abstract}
 We present a detailed study of the spherically symmetric solutions in Lorentz breaking massive gravity. There is an undetermined function $\mathcal{F}(X, w_1, w_2, w_3)$ in the action of St\"{u}ckelberg fields $S_{\phi}=\Lambda^4\int{d^4x\sqrt{-g}\mathcal{F}}$, which should be resolved through physical means. In the general relativity, the spherically symmetric solution to the Einstein equation is a benchmark and its massive deformation also play a crucial role in Lorentz breaking massive gravity. $\mathcal{F}$ will satisfy the constraint equation $T_0^1=0$ from the spherically symmetric Einstein tensor $G_0^1=0$, if we maintain that any reasonable physical theory should possess the spherically symmetric solutions.  The St\"{u}ckelberg field $\phi^i$ is taken as a 'hedgehog' configuration $\phi^i=\phi(r)x^i/r$, whose stability is guaranteed by the topological one. Under this ans\"{a}tz, $T_0^1=0$ is reduced to $d\mathcal{F}=0$. The functions $\mathcal{F}$ for $d\mathcal{F}=0$ form a commutative ring $R^{\mathcal{F}}$. We obtain an expression of solution to the functional differential equation with spherically symmetry if $\mathcal{F}\in R^{\mathcal{F}}$. If $\mathcal{F}\in R^{\mathcal{F}}$ and $\partial\mathcal{F}/\partial X=0$, the functions $\mathcal{F}$ form a subring $S^{\mathcal{F}}\subset R^{\mathcal{F}}$. We show that the metric is Schwarzschild, Schwarzschild-AdS or Schwarzschild-dS if $\mathcal{F}\in S^{\mathcal{F}}$. When $\mathcal{F}\in R^{\mathcal{F}}$ but $\mathcal{F}\notin S^{\mathcal{F}}$, we will obtain some new metric solutions, including the furry black hole and beyond. Using the general formula and the basic property of function ring $R^{\mathcal{F}}$, we give some analytical examples and their phenomenological applications. Furthermore, we also discuss the stability of gravitational field by the analysis of Komar integral and the results of QNMs.
\end{abstract}

%
%
%
%
%

\section{Introduction}

It is an interesting question whether general relativity (GR) is a solitary theory from both theoretical and phenomenological sides. One of modifying gravity theories is the massive deformation of GR. A comprehensive review of massive gravity can be found in \cite{Rham}. We can divide the massive gravity theories into two varieties: Lorentz invariant type (LI) and Lorentz breaking type (LB). LB massive gravity can be physically viable, which is free from pathologies such as the ghosts, the van Dam-Veltman-Zakharov (vDVZ) discontinuity due to extra graviton polarization and strong coupling at the low energy scale (see e. g., Refs. \cite{Rubakov, Dubovsky1} and references therein). In most of LB cases the energy cutoff $(mM_{pl})^{\frac{1}{2}}$ is quite high in comparison with that in LI cases, such that LB massive gravity can avoid some phenomenological shortcomings from LI massive gravity \cite{Comelli1}. For instance, there is a class of LB massive gravity models in which the solar system constraints are satisfied for larger graviton mass \cite{Dubovsky2}. The relic of gravitational waves engendered during inflation epoch may constitute today the cold dark matter in the Universe and would bring to a distinguishingly monochromatic signal in the gravitational wave detectors \cite{Dubovsky3}. However, the significant contribution of massive gravitons to the local dark halo density have ruled out essentially \cite{Pshirkov}. Recently, the massive gravity theories with five propagating degrees of freedom were made in a series of paper \cite{Comelli2, Comelli3, Comelli4}. The Lorentz breaking in their gravitational sector is inescapable to concede a nontrivial cosmology for a spatially flat Universe \cite{Comelli1}.

In GR, the spherically symmetric vacuum solution to the Einstein equation is a benchmark and its massive deformation also play a crucial role in LB massive gravity. The exact spherically symmetric solutions in some massive gravity models show a variety of different features \cite{Bebronne1, Dubovsky4, Bebronne2, Comelli5}. They differ from the Schwarzschild solution, which depend on two parameters: the mass $M$ and the scalar charge $S$. Especially, the solutions show a nonanalytic hair in the form of a powerlike term $r^{-\lambda}$ \cite{Bebronne1}. The solutions may possess a horizon depending upon the parameters $M$ and $S$, which become candidates of black holes.

In most theories of LB massive gravity, the Lorentz group is broken down to a $SO(3)$ rotation group. A class of promising LB massive gravity theories is especially interesting where the ghosts are absent because there is the residual symmetry $x^i\rightarrow x^i+\xi^i(t)$ on flat space for three arbitrary functions $\xi^i(t)$. In the picture of St\"{u}ckelberg fields $\phi^0$, $\phi^i$ ($i=1, 2, 3$), this corresponds with the internal symmetry $\phi^i\rightarrow\phi^i+\xi^i(\phi)$. Thus, the mass term should be a function $\mathcal{F}$ of the dimensionless quantities $X=g^{\mu\nu}\partial_{\mu}\phi^0\partial_{\nu}\phi^0$ and $W^{ij}=(g^{\mu\nu}-X^{-1}\partial^{\mu}\phi^0\partial^{\nu}\phi^0)\partial_\mu\phi^i\partial_\nu\phi^j$ \cite{Rubakov}. The authors of Refs. \cite{Bebronne1, Comelli5} studied the black hole solution of models described by the designated function $\mathcal{F}$.

In this work, we release from the limitations of designated function $\mathcal{F}$ so that the Einstein equations become the functional differential equations at present. We have found the functional formula of spherically symmetric solutions. In GR, the Birkhoff's theorem is the statement that the Schwarzschild metric is the unique vacuum solution with spherical symmetry. The standard Birkhoff's theorem cannot be applied because there are the St\"{u}ckelberg fields in the massive gravity. Therefore, the function $\mathcal{F}$ must be not arbitrary, but restrictive if we consider a static spherically symmetric system. The St\"{u}ckelberg field $\phi^i$ is taken as a 'hedgehog' configuration, whose topological property guarantees that the configuration is stable. The proper functions $\mathcal{F}$ should satisfy $d\mathcal{F}=0$ from the condition of spherically symmetric solution $T_0^1=0$ where $T_\mu^\nu$ is the energy-momentum tensor of the  St\"{u}ckelberg fields. Under the hedgehog ans\"{a}tz, the functions $\mathcal{F}$ form a commutative ring $R^{\mathcal{F}}$. The basic property of $R^{\mathcal{F}}$ affords us how to choose $\mathcal{F}$ for the various solutions. Before leading to new solutions, we first checked a well-known example using our universal formula of the spherically symmetric solutions. The check has a double significance, which not only checked the known solution but also checked the formula itself. If the functions $\mathcal{F}\in R^{\mathcal{F}}$ and $\partial\mathcal{F}/\partial{X}=0$, $\mathcal{F}$ form a subring $S^{\mathcal{F}}\subset R^{\mathcal{F}}$. We find that the metric is Schwarzschild, Sch-AdS or Sch-dS if $\mathcal{F}\in S^{\mathcal{F}}$. We obtain new analytical solutions as well as the well-known example when $\mathcal{F}\in R^{\mathcal{F}}$ but $\mathcal{F}\notin S^{\mathcal{F}}$. Especially, one of new solutions has interesting astrophysics applications.

The paper is organized as follows: in section 1 we introduce the Lorentz breaking massive gravity. In section 2 we shall answer how to choose $\mathcal{F}$ and show that the proper functions $\mathcal{F}$ form a commutative ring $\mathcal{R}^{\mathcal{F}}$. Furthermore, we shall give the functional differential equations and the expression of static spherically symmetric solutions. Then, we shall give some analytical solutions, classification of the functions $\mathcal{F}$, and discuss stability of the St\"{u}ckelberg field configuration in section 3, analyse the stability of gravitational field of these solutions in section 4 and study the phenomenological consequences for new solution in section 5. Finally, section 6 is devoted to the conclusion and discussion.

\section{Static spherically symmetric equations and solutions}
In this work, we discussed the spherically symmetric solutions in a wide class of promising LB massive gravity theories, where the only propagating degree of freedom is a single graviton \cite{Dubovsky1}. Following St\"{u}ckelberg \cite{Stuckelberg}, general covariance is restored by a set of Goldstone fields, which are nonlinearly interacting with the metric field. The action can be written as $S=S_{GR}+S_{\phi}$, and
\begin{equation}\label{1.1}
S_{\phi}=\Lambda^4\int{d^4x\sqrt{-g}\mathcal{F}(X,W^{ij})}.
\end{equation}
$S_{\phi}$ describes a set of four St\"{u}ckelberg fields $\phi^0$, $\phi^{i}$ ($i=1, 2, 3$) transforming under a diffeomorphism (diff) $\delta x^\mu=\xi^\mu(x)$ as simple scalars. One can structure diff invariant function $\mathcal{F}$ using dimensionless quantities $X=g^{\mu\nu}\partial_\mu\phi^0\partial_\nu\phi^0$ and $W^{ij}=(g^{\mu\nu}-X^{-1}\partial^\mu\phi^0\partial^\nu\phi^0)\partial_\mu\phi^i\partial_\nu\phi^j$. The fields $\phi^0$ and $\phi^i$ have spacetime dependent vacuum expectation values
\begin{eqnarray}\label{1.2}
g_{\mu\nu}=\eta_{\mu\nu}, \phi^0=at, \phi^i=bx^i.
\end{eqnarray}
If a function $\mathcal{F}$ is chosen, we can redefine fields $\phi^0$ and $\phi^i$ so that $a=b=1$. The action (\ref{1.1}) is invariant under $\phi^i\rightarrow\phi^i+\xi^i(\phi^0)$, where $\xi^i$ are arbitrary functions of $\phi^0$. This symmetry turns out that the perturbations about the vacuum solution (\ref{1.2}) are nonpathological \cite{Dubovsky1}. One requires that the background breaks only the invariance under boosts but not under the whole Lorentz group. Then, the function $\mathcal{F}$ depends on $W^{ij}$ through three combinations $w_n=TrW^n$ ($n=1,2,3$) which is necessary for the background to be invariant under the $SO(3)$ symmetry in the $\phi^i$ internal space \cite{Bebronne1, Comelli5}. The constant $\Lambda$ has the dimension of mass, and $S_\phi$ is understood as the low-energy effective action valid below the scale $\Lambda$.

\subsection{How do we choose $\mathcal{F}$?}
The static spherically symmetric ans\"{a}tz can be written as follows \cite{Bebronne1}
\begin{eqnarray}\label{2.1}
ds^2&=&\alpha(r)dt^2-\beta(r)dr^2-r^2d\Omega^2,\nonumber\\
\phi^0&=&t+h(r),\nonumber\\
\phi^i&=&\frac{1}{r}\phi(r)x^i,
\end{eqnarray}
where $\phi^i$ is the so-called 'hedgehog' configuration. We note that the ans\"{a}tz (\ref{2.1}) is not the most general one consistent with the symmetries. The spatial St\"{u}ckelberg field is $\phi^i=bx^i$ in the end (see Lemma 1 in Appendix A). In other words, they are actually fixed to their unitary values. However it may be that the static spherically symmetric ans\"{a}tz is written in different coordinates in unitary gauge \cite{berezhian}.

This configuration contains two additional radial functions, $h(r)$ and $\phi(r)$ as compared with GR. Since $h(r)=constant$ corresponds to vacuum solution, we might as well assume $h'(r)\neq0$ for nontrivial solutions. Using (\ref{2.1}), we can express $w_n$ and $X$ by the following relations
\begin{eqnarray}\label{2.2}
w_1&=&-\phi'^2(\alpha\beta)^{-1}X^{-1}-2\phi^2r^{-2},\nonumber\\
w_2&=&\phi'^4(\alpha\beta)^{-2}X^{-2}+2\phi^4r^{-4},\nonumber\\
w_3&=&-\phi'^6(\alpha\beta)^{-3}X^{-3}-2\phi^6r^{-6},
\end{eqnarray}
and
\begin{equation}\label{2.3}
X=\frac{\beta-\alpha h'^2}{\alpha\beta}.
\end{equation}
The non-zero components of energy-momentum tensor $T_\mu^\nu$ for the Goldstone fields are given by the following expressions
\begin{eqnarray}
T_0^0&=&\Lambda^4[\mathcal{F}+2(\alpha\beta X)^{-1}\phi'^2\mathcal{F}_1-4(\alpha\beta X)^{-2}\phi'^4\mathcal{F}_2+6(\alpha\beta X)^{-3}\phi'^6\mathcal{F}_3],\label{2.4}\\
T_0^1&=&-\frac{2\Lambda^4h'}{\beta}[\mathcal{F}_X+(\alpha\beta)^{-1}X^{-2}\phi'^2\mathcal{F}_1-2(\alpha\beta)^{-2}X^{-3}\phi'^4\mathcal{F}_2+3(\alpha\beta)^{-3} X^{-4}\phi'^6\mathcal{F}_3],\nonumber\\\label{2.5}\\
T_1^1&=&\Lambda^4[\mathcal{F}-2\mathcal{F}_X X+\alpha^{-1}(2\mathcal{F}_X+2(\alpha\beta)^{-1} X^{-2}\phi'^2\mathcal{F}_1-4(\alpha\beta)^{-2}X^{-3}\phi'^4\mathcal{F}_2\nonumber\\&+&6(\alpha\beta)^{-3} X^{-4}\phi'^6\mathcal{F}_3)],\label{2.6}\\
T_2^2&=&T_3^3=\mathcal{F}+2\frac{\phi^2}{r^2}\mathcal{F}_1-4\frac{\phi^4}{r^4}\mathcal{F}_2+6\frac{\phi^6}{r^4}\mathcal{F}_3,\label{2.7}
\end{eqnarray}
where $\mathcal{F}_i\equiv\partial\mathcal{F}/\partial w_i$ and $\mathcal{F}_X\equiv\partial\mathcal{F}/\partial X$. The non-zero components of the Einstein tensor are
\begin{eqnarray}
G_0^0&=&\frac{1}{r^2}[1-(\frac{r}{\beta})'],\label{2.8}\\
G_1^1&=&\frac{1}{r^2}(1-\frac{\alpha+r\alpha'}{\alpha\beta}),\label{2.9}\\
G_2^2&=&G_3^3=-\frac{1}{4r}[\frac{\alpha'+r\alpha''}{\alpha\beta}+(\frac{2\alpha+r\alpha'}{\alpha\beta})'].\label{2.10}
\end{eqnarray}
From the Einstein equations $G_\mu^\nu=M_{pl}^{-2}T_\mu^\nu$ and $G_0^1=0$ we have $T_0^1=0$. It is easy to prove that $T_0^0-T_1^1$ is proportional to $T_0^1$ and $T_0^0=T_1^1$ implies $G_0^0=G_1^1$. Thus, we have $\alpha(r)\beta(r)=1$, and the Einstein equations can be rewritten as
\begin{eqnarray}
\frac{\alpha'}{r}+\frac{\alpha-1}{r^2}-\frac{m^2}{2}(\mathcal{F}-2X\mathcal{F}_X)=0,\label{11}\\
\frac{\alpha''}{2}+\frac{\alpha'}{r}-\frac{m^2}{2}(\mathcal{F}+\frac{2\phi^2}{r^2}\mathcal{F}_1-\frac{4\phi^4}{r^4}\mathcal{F}_2+\frac{6\phi^6}{r^6}\mathcal{F}_3)=0,\label{12}\\
\mathcal{F}_X+\frac{\phi'^2}{X^2}\mathcal{F}_1-2\frac{\phi'^4}{X^3}\mathcal{F}_2+3\frac{\phi'^6}{X^4}\mathcal{F}_3=0.\label{13}
\end{eqnarray}
From $\alpha\beta=1$, (\ref{2.2}) and (\ref{2.3})can be reduced to
\begin{eqnarray}\label{2.11}
w_1=-(\phi'^2X^{-1}+2\phi^2r^{-2}),\nonumber\\
w_2=(\phi'^2X^{-1})^2+2(\phi^2r^{-2})^2,\nonumber\\
w_3=-[(\phi'^2X^{-1})^3+2(\phi^2r^{-2})^3],
\end{eqnarray}
and
\begin{equation}\label{17}
X=\frac{1-\alpha^2h'^2}{\alpha}.
\end{equation}

Eqs. (\ref{11})-(\ref{13}) are the system containing three differential equations if and only if we take a fixed function $\mathcal{F}$. From this point of view, the system contains three equations for three unknowns, $\alpha$, $\phi$ and $X$. Even though it is a highly non-linear coupling system, it is solvable. Our purpose is to find an expression of spherically symmetric solution and analytical solutions for some particular choices of the function $\mathcal{F}$. We adopt a strategy to achieve this purpose as follows: The first step is that (\ref{13}) is reduced to an algebraic equation if we assume $\mathcal{F}$ is a polynomial of $X^{-1}$, $w_1$, $w_2$ and $w_3$. Thus, we show that there is only solution $\phi(r)=br$ and (\ref{11})-(\ref{13}) become a system containing two functional differential equations. Secondly, we find an expression of $\alpha(r)$ for any polynomial $\mathcal{F}$. Thirdly, we obtain $X$ using the relation between $\alpha$ and $X$. Finally, we obtain $\phi^0$ using the relation between $X$, $\alpha$ and $\phi^0$.

To solve the functional-differential equation (\ref{13}), we realize that only one of variables $X, w_1, w_2$ and $w_3$ is effective. This occurs if there are subsidiary relationships constraining all the variables to have values dependent on the value of $X$. These relationships may be represented by equations
\begin{eqnarray}\label{18}
w_i=w_i(X), (i=1, 2, 3)
\end{eqnarray}
There is only solution for (\ref{18}),
\begin{equation}\label{19}
\phi(r)=br,
\end{equation}
where $b$ is an undetermined constant and it will be used in (\ref{2.11}). Fortunately, we can prove that (\ref{19}) is the only solution to Einstein equations (\ref{11})-(\ref{13}) under the hedgehog ans\"{a}tz (\ref{2.1}) and $\mathcal{F}$ is a polynomial of $X^{-1}$, $w_1$, $w_2$ and $w_3$ (see Lemma 1 in Appendix A). Thus, (\ref{13}) can be rewritten as
\begin{eqnarray}\label{20}
d\mathcal{F}=\mathcal{F}_XdX+\mathcal{F}_1dw_1+\mathcal{F}_2dw_2+\mathcal{F}_3dw_3=0.
\end{eqnarray}

We note that the gravitational fields obey the Bianchi identity $\nabla_\mu G^{\mu\nu}=0$ at present which implies the conservational requirement $\nabla_\mu T^{\mu\nu}=0$. The energy-momentum tensor is really conservation in the $\phi=br$, so there is consistency between the Bianchi identity and the additional restriction (\ref{13}) on $\mathcal{F}$.

We consider $\mathcal{F}$ is a polynomial below,
\begin{equation}\label{2.13}
\mathcal{F}=\sum_{l_0,\cdots, l_3}a_{l_0l_1l_2l_3}(\frac{1}{X})^{l_0}w_1^{l_1}w_2^{l_2}w_3^{l_3}
\end{equation}
where $l_0,\cdots, l_3$ are positive integers. Mathematically, a commutative ring $R^{\mathcal{F}}$ is a set with two laws of composition $+$ and $\times$, called addition and multiplication, that satisfies rules: (i) With the law of composition $+$, $R^{\mathcal{F}}$ is an abelian group that we denote by $R^{\mathcal{F}^+}$; its identity is denoted by 0; (ii) Multiplication is commutative and associative, and has an identity denoted by 1; (iii) With the law of distribution $(f^i+f^j)f^k=f^if^k+f^jf^k$ and $f^k(f^i+f^j)=f^kf^i+f^kf^j$ for all $f^i$, $f^j$, $f^k$ in $R^{\mathcal{F}}$. In our case, all solutions (\ref{2.13}) of $d\mathcal{F}=0$ form a function ring $R^{\mathcal{F}}$. Remarkably, there exist different solutions for $\alpha(r)$ and $h(r)$, even though $\phi^i(r)$ is provided with same form. The algebraic feature of $R^{\mathcal{F}}$ shall aid us to find new solutions $\alpha(r)$ and $h(r)$. Especially, all solutions $\mathcal{F}=\sum_{l_1,\cdots, l_3}a_{0l_1l_2l_3}w_1^{l_1}w_2^{l_2}w_3^{l_3}$ form a subring $S^{\mathcal{F}}\subset R^{\mathcal{F}}$. If $\mathcal{F}\in S^{\mathcal{F}}$, we will prove that the solutions of Einstein equation are the Schwarzschild, Sch-AdS or Sch-dS respectively.

\subsection{Functional differential equations}

Under the hedgehog ans\"{a}tz (\ref{2.1}), there is the only solution $\phi=br$ for Einstein equations (\ref{11})-(\ref{13}). Thus we choose $\mathcal{F}\in R^{\mathcal{F}}$, the Einstein equations are reduced to the following functional differential equations
\begin{eqnarray}
\frac{\alpha'}{r}&+&\frac{\alpha-1}{r^2}-\frac{m^2}{2}\mathcal{K}(\mathcal{F})=0,\label{2.20}\\
\frac{\alpha''}{2}&+&\frac{\alpha'}{r}-\frac{m^2}{2}\mathcal{H}(\mathcal{F})=0,\label{2.21}
\end{eqnarray}
and $\alpha\beta=1$. Here $m^2=2\Lambda^4/M_{pl}^2$ and the functionals $\mathcal{K}(\mathcal{F})$ and $\mathcal{H}(\mathcal{F})$ are defined as
\begin{eqnarray}
\mathcal{K}(\mathcal{F})&=&\mathcal{F}-2X\mathcal{F}_X,\label{2.22}\\
\mathcal{H}(\mathcal{F})&=&\mathcal{F}+2b^2\mathcal{F}_1-4b^4\mathcal{F}_2+6b^6\mathcal{F}_3.\label{2.23}
\end{eqnarray}
(\ref{2.20}) and (\ref{2.21}) are well defined. Since $\mathcal{F}$ is a function of $X$, $w_1$, $w_2$ and $w_3$, the functionals $\mathcal{K}(\mathcal{F})$ and $\mathcal{H}(\mathcal{F})$ are also ones. Furthermore, $w_1$, $w_2$ and $w_3$ are the polynomials of $X^{-1}$ from (\ref{2.11}) and (\ref{19}), so $\mathcal{K}$ and $\mathcal{H}$ are also looked upon as functions of $X$. Thus, $X$ can be expressed as a function of $\alpha$, $\alpha'$ and $r$ from (\ref{2.20}), and (\ref{2.21}) becomes an ordinary differential equation of $\alpha$. Therefore, we can obtain $\alpha$ from (\ref{2.21}), then we have the solution $\phi^0$ from (\ref{17}) and (\ref{2.20}).

We solve formally (\ref{2.20}) as follows
\begin{equation}\label{2.24}
X=\mathcal{K}^{-1}(y),
\end{equation}
where
\begin{equation}\label{2.25}
y=\frac{2}{m^2}(\frac{\alpha'}{r}+\frac{\alpha-1}{r^2}).
\end{equation}
According to the fundamental theorem of algebra, every nonconstant polynomial has a complex root. However, if $\mathcal{F}\in R^{\mathcal{F}}$, $\mathcal{F}+c$ is also in $R^{\mathcal{F}}$ so that we always obtain real $\mathcal{K}^{-1}(y)$. This result can be generalized for any differentiable function in $R^{\mathcal{F}}$.

There is a mathematical identity relation
\begin{equation}\label{2.26}
2\alpha''+4r^{-1}\alpha'=m^2ry'+2m^2y,
\end{equation}
which is the key to the solving process. Using (\ref{2.26}), (\ref{2.21}) can be rewritten as
\begin{equation}\label{2.27}
\frac{r}{2}y'+y-\mathcal{H}(\mathcal{K}^{-1}(y))=0,
\end{equation}
where $\mathcal{H}(\mathcal{K}^{-1}(y))$ is also the functional of $\mathcal{F}$, and we can regard it as a function of $y$. Therefore, (\ref{2.27}) is a separable first-order equation of $y$ in reality. When we solve (\ref{2.27}), there are two cases: (i) $\mathcal{H}=\mathcal{K}$; (ii)$\mathcal{H}\neq\mathcal{K}$. We have $\mathcal{F}_X=0$ if $\mathcal{H}(\mathcal{F})=\mathcal{K}(\mathcal{F})$ (see Lemma 2 in Appendix A).

\subsection{The Schwarzschild, AdS and dS solutions}
If $\mathcal{F}\in S^{\mathcal{F}}$, we have $\mathcal{H}=\mathcal{K}$. In this case, (\ref{2.27}) has only constant solutions, $y=const.$, which correspond the Schwarzschild, Sch-AdS or Sch-dS solutions respectively from (\ref{2.25}). That is to say
\begin{equation}\label{2.28}
\alpha =
\cases{
1-\frac{r_s}{r}&for $y=0$,\\
1-\frac{r_s}{r}\pm\mu^2r^2&for $y=\pm\frac{6\mu^2}{m^2}$.}
\end{equation}
From (\ref{2.3}) for the Schwarzschild solution, we have
\begin{equation}\label{2.29}
\fl{h(r) =
\cases{
A+r_s\cosh^{-1}{B}+\frac{(2-\mathcal{K}^{-1}(0))r_s}{2C}\ln{(A+Cr+\frac{\mathcal{K}^{-1}(0)r_s}{2C})}+u_0 & for $\mathcal{K}^{-1}(0)<1$\\
r+r_s\ln{(r-r_s)}+u_0&for $\mathcal{K}^{-1}(0)=0$\\
A+r_s\ln{(\frac{\sqrt{r}-\sqrt{r_s}}{\sqrt{r}+\sqrt{r_s}})}+u_0&for $\mathcal{K}^{-1}(0)=1$}}
\end{equation}
where $u_0$ is an integral constant and
\begin{eqnarray}\label{2.30}
A&=&[(1-\mathcal{K}^{-1}(0))r^2+\mathcal{K}^{-1}(0)r_s r]^{\frac{1}{2}},\nonumber\\
B&=&\frac{(2-\mathcal{K}^{-1}(0))r_s r+\mathcal{K}^{-1}(0)r_s^2}{|\mathcal{K}^{-1}(0)|r_s(r-r_s)},\nonumber\\
C&=&[1-\mathcal{K}^{-1}(0)]^{\frac{1}{2}}.
\end{eqnarray}
It is worth to note that $\phi^0(r)$ is not real solution if $\mathcal{K}^{-1}(0)>1$.

In the case of dS solution $\alpha=1-\mu^2r^2$, we require $\mathcal{K}^{-1}(-\frac{6\mu^2}{m^2})\leq1$ for $\phi^0(r)$ being real solution. From (\ref{2.3}), we have
\begin{equation}\label{2.31}
\fl{h(r) =
\cases{
-\frac{1}{2\mu}\ln{(1-\mu^2r^2)}+u_0& for $\mathcal{K}^{-1}(-\frac{6\mu^2}{m^2})=1$\\
\frac{1}{2\mu}\ln{[\frac{A(1-B+C)}{B(1+A+C)}(\frac{C-D}{C+D})^{\sqrt{\mathcal{K}^{-1}(-\frac{6\mu^2}{m^2})}}]}+u_0&for $0<\mathcal{K}^{-1}(-\frac{6\mu^2}{m^2})<1$\\
\frac{1}{2\mu}\ln{\frac{A}{B}}+u_0&for $\mathcal{K}^{-1}(-\frac{6\mu^2}{m^2})=0$\\
\frac{1}{2\mu}[\ln{\frac{A(1-B+C)}{B(1+A+C)}}-2\sqrt{-\mathcal{K}^{-1}(-\frac{6\mu^2}{m^2})}\sin{E}]+u_0&for $\mathcal{K}^{-1}(-\frac{6\mu^2}{m^2})<0$}}
\end{equation}
where $u_0$ is an integral constant and
\begin{eqnarray}\label{2.35}
A&=&1+\mu r,\nonumber\\
B&=&1-\mu r,\nonumber\\
C&=&[1-\mathcal{K}^{-1}(-\frac{6\mu^2}{m^2})(1-\mu r^2)]^{\frac{1}{2}},\nonumber\\
D&=&\sqrt{\mathcal{K}^{-1}(-\frac{6\mu^2}{m^2})}\mu r,\nonumber\\
E&=&\sqrt{\frac{\mathcal{K}^{-1}(-\frac{6\mu^2}{m^2})}{\mathcal{K}^{-1}(-\frac{6\mu^2}{m^2})-1}}\mu r.
\end{eqnarray}

For the case of AdS solution $\alpha=1+\mu^2r^2$, $\phi^0(r)$ is not real as $\mathcal{K}^{-1}(\frac{6\mu^2}{m^2})>0$. From (\ref{2.3}), we obtain
\begin{equation}\label{2.36}
\fl{h(r)=
\cases{
\frac{1}{\mu}\tan^{-1}{\mu r}+u_0&for $\mathcal{K}^{-1}(\frac{6\mu^2}{m^2})=0$\\
\frac{1}{\mu}\tan^{-1}{\frac{\mu r}{\sqrt{1-A}}}+\frac{\sqrt{-\mathcal{K}^{-1}(\frac{6\mu^2}{m^2})}}{\mu}\ln{(\sqrt{\mathcal{K}^{-1}(\frac{6\mu^2}{m^2})(A-1)}-B)}+u_0&for $\mathcal{K}^{-1}(\frac{6\mu^2}{m^2})<0$}}
\end{equation}
where $u_0$ is an integral constant and
\begin{eqnarray}\label{2.38}
A&=&[\mathcal{K}^{-1}(\frac{6\mu^2}{m^2})](1+\mu^2r^2),\nonumber\\
B&=&[\mathcal{K}^{-1}(\frac{6\mu^2}{m^2})]\mu r.
\end{eqnarray}

Similarly, we can obtain $h(r)$ for Sch-AdS or Sch-dS from (\ref{2.3}). In these cases, the metric is independent with the integral constant $u_0$, so it only represents a residual freedom for the St\"{u}ckelberg field $\phi^0$.

\subsection{The expression of solution in the case of $\mathcal{H}\neq\mathcal{K}$}
If $\mathcal{F}\in R^{\mathcal{F}}$ but $\mathcal{F}\notin S^{\mathcal{F}}$, we have $\mathcal{H}\neq\mathcal{K}$. In this case, we obtain the formal integration from (\ref{2.27}) as follows
\begin{equation}\label{2.39}
\frac{r^2}{r_0^2}=\exp{[\int_{y_0}^{y}\frac{dy}{\mathcal{H}(\mathcal{K}^{-1}(y))-y}]}.
\end{equation}
Thus we expressed $y$ as a function of $r$ using the corresponding relation (\ref{2.39}) between $r$ and $y$, i.e., $y=y(r,u_0)$ where $u_0$ is an integral constant. (\ref{2.25}) can be rewritten as
\begin{equation}\label{2.40}
\frac{\alpha'}{r}+\frac{\alpha-1}{r^2}=\frac{m^2}{2}y(r,u_0),
\end{equation}
then we have the formula of ordinary solution
\begin{equation}\label{2.41}
\alpha=1+\frac{m^2}{2r}\int{r^2y(r,u_0)dr}.
\end{equation}

From (\ref{2.3}), (\ref{2.20}) and $\alpha\beta=1$, we obtain
\begin{equation}\label{2.42}
h(r)=\pm\int{\frac{dr}{\alpha}[1-\alpha\mathcal{K}^{-1}(y)]^{\frac{1}{2}}}.
\end{equation}
Thus, we finally obtain the expression of solution for the functional differential equations in the $\mathcal{H}\neq\mathcal{K}$ case,
\begin{eqnarray}\label{2.43}
\alpha&=&\beta^{-1}=1+\frac{m^2}{2r}\int{r^2y(r,u_0)dr},\nonumber\\
\phi^0&=&t\pm\int{\frac{dr}{\alpha}[1-\alpha\mathcal{K}^{-1}(y)]^{\frac{1}{2}}},\nonumber\\
\phi^i&=&bx^i.
\end{eqnarray}

\section{Analytical solutions beyond known solutions in GR}
\subsection{The solutions of furry black holes}
To find new solution beyond Schwarzschild, AdS and dS, we must take $\mathcal{F}\in R^{\mathcal{F}}$ but $\mathcal{F}\notin S^{\mathcal{F}}$. In the simplest case, we choose that $\mathcal{F}$ satisfies $\mathcal{K}(\mathcal{F})=c\mathcal{H}(\mathcal{F})$, $c\in\mathcal{R}$.
Thus, we have
\begin{equation}\label{3.1}
\mathcal{H}(\mathcal{K}^{-1}(y))=cy,
\end{equation}
and (\ref{2.27}) is reduced to
\begin{equation}\label{3.2}
\frac{r}{2}y'+(1-c)y=0.
\end{equation}
The ordinary solution of (\ref{2.27}) is
\begin{equation}\label{3.3}
y=u_0r^{2(c-1)},
\end{equation}
where $u_0$ is integral constant. Substituting (\ref{3.1}) and (\ref{3.3}) into (\ref{2.43}), we obtain the solution
\begin{eqnarray}\label{3.4}
\alpha&=&\beta^{-1}=1-\frac{r_s}{r}-\frac{S}{r^{\lambda}},\nonumber\\
\phi^0&=&t\pm\int{\frac{dr}{\alpha}[1-\alpha\mathcal{K}^{-1}(y)]^{\frac{1}{2}}},\nonumber\\
\phi^i&=&bx^i,
\end{eqnarray}
where $\lambda=-2c$, $S=-\frac{m^2u_0}{4c+2}$ and $r_s$ is another integral constant.

Next, we reconsider a well-known example \cite{Bebronne1} to check up the formula (\ref{3.4}). Taking the function
\begin{equation}\label{3.5}
\mathcal{F}=c_0(\frac{b^2}{X}+w_1)+c_1(w_1^3-3w_1w_2-6b^4w_1+2w_3-12b^6),
\end{equation}
where $c_0$ and $c_1$ are dimensionless constants. Two additional conditions should be enforced on these two constants: (i) Non-tachyonic condition $c_0-6c_1>0$; (ii) No ghost condition $c_0>0$. Thus, we have
\begin{eqnarray}\label{3.6}
\mathcal{K}^{-1}(y)=\frac{2c_0b^2}{y+2c_0b^2}\quad\textmd{and}\quad c=\frac{6c_1b^4}{c_0}.
\end{eqnarray}
Substituting (\ref{3.3}) and (\ref{3.6}) into (\ref{3.4}), we obtain again the solution in \cite{Bebronne1} by the markedly dissimilar way,
\begin{eqnarray}\label{3.7}
\alpha&=&\beta^{-1}=1-\frac{r_s}{r}-\frac{S}{r^{\lambda}},\nonumber\\
\phi^0&=&t\pm\int{\frac{dr}{\alpha}[1-\alpha(1+\frac{u_0}{2c_0b^2}r^{2(c-1)})^{-1}]^{\frac{1}{2}}},\nonumber\\
\phi^i&=&bx^i.
\end{eqnarray}

We note that the dimension of $u_0$ is $[Length]^{\lambda+2}$ in the furry black hole (\ref{3.7}). The parameter $u_0$ satisfies $u_0\geq-\frac{r_s^2M_{pl}^2}{4\Lambda^4}$ if we take $c_1=-\frac{c_0}{6b^4}$. For this black hole, we have $u_0=-\frac{3r_s^2M_{pl}^2}{16\Lambda^4}$ if it has solar mass and horizon $r_h=\frac{3r_s}{4}$.

By the straightforward calculation, we find a more simple solution $\mathcal{F}(X, w_1, w_2)$ which satisfies $\mathcal{K}(\mathcal{F})=c\mathcal{H}(\mathcal{F})$ as follows
\begin{equation}\label{3.8}
\mathcal{F}=c_0(\frac{b^2}{X}+w_1)+c_1(w_1^2-w_2+4b^2w_1+6b^4),
\end{equation}
and
\begin{equation}\label{3.9}
c=-\frac{2c_1b^2}{c_0},
\end{equation}
where dimensionless constants $c_0$ and $c_1$ should satisfy the constraints $c_0+4c_1>0$ and $c_0>0$.

Thus, we have
\begin{equation}\label{3.11}
\mathcal{K}^{-1}(y)=\frac{2c_0b^2}{y+2c_0b^2},
\end{equation}
and we finally obtain the solution
\begin{eqnarray}\label{3.12}
\alpha&=&\beta^{-1}=1-\frac{r_s}{r}-\frac{S}{r^{\lambda}},\nonumber\\
\phi^0&=&t\pm\int{\frac{dr}{\alpha}[1-\alpha(1+\frac{u_0}{2c_0b^2}r^{2(c-1)})^{-1}]^{\frac{1}{2}}},\nonumber\\
\phi^i&=&bx^i.
\end{eqnarray}

Using the same procedure, we take
\begin{equation}\label{3.13}
\mathcal{F}=c_0(\frac{b^6}{X^3}+w_3)+c_1(w_1^2+3b^2w_1+3b^4)(w_1^2-w_2+4b^2w_1+6b^4),
\end{equation}
and the analytical solution is
\begin{eqnarray}\label{3.14}
\alpha&=&\beta^{-1}=1-\frac{r_s}{r}-\frac{S}{r^{\lambda}},\nonumber\\
\phi^0&=&t\pm\int{\frac{dr}{\alpha}[1-\alpha(1+\frac{u_0}{6c_0b^6}r^{-\lambda-2})^{-\frac{1}{3}}]^{\frac{1}{2}}},\nonumber\\
\phi^i&=&bx^i,
\end{eqnarray}
where $S=\frac{m^2u_0}{2\lambda-2}$, $\lambda=\frac{4c_1b^2}{3c_0}$ and $u_0$, $r_s$ are integral constants.

There are certainly the infinite solutions with $\mathcal{F}\in\mathcal{R}^{\mathcal{F}}$ and $\mathcal{K}(\mathcal{F})=c\mathcal{H}(\mathcal{F})$. Furthermore, each fixed $c$ can generate a subgroup $S_c^{\mathcal{F}}$ which satisfies $\mathcal{K}(\mathcal{F})=c\mathcal{H}(\mathcal{F})$ if $\mathcal{F}\in S_c^{\mathcal{F}}$ (see Lemma 3 in Appendix A). For any $\mathcal{F}$ that satisfies $\mathcal{K}(\mathcal{F})=c\mathcal{H}(\mathcal{F})$, we have the same form for metric
\begin{equation}\label{3.15}
\alpha=\beta^{-1}=1-\frac{r_s}{r}-\frac{S}{r^{\lambda}}.
\end{equation}
This metric differ from the Schwarzschild solution by an additional powerlike term $r^{-\lambda}$. Such solution may possess an event horizon depending upon the parameters of the function $\mathcal{F}$, which become candidates of modified black hole. In other words, such black holes can be described by two physical parameters: Schwarzschild radius $r_s$ and scalar charges $S$, so that they are dubbed furry black holes \cite{Comelli5}.

\subsection{An analytical solution beyond furry black holes}
If we want find the solutions beyond furry black holes, we have to choose $\mathcal{F}\in\mathcal{R}^{\mathcal{F}}$ and $\mathcal{K}(\mathcal{F})\neq c\mathcal{H}(\mathcal{F})$. Using the expression of solution (\ref{2.43}), we are able to obtain new solution. For instance, we take $\mathcal{F}$ as follows
\begin{equation}\label{3.16}
\mathcal{F}=c_0(\frac{b^2}{X}+w_1+2b^2)+c_1(w_1+b^2)(w_1^2-w_2+4b^2w_1+6b^4),
\end{equation}
where $c_0$ and $c_1$ are dimensionless constants, and $b^2=(c_0/2c_1)^{\frac{1}{2}}$. Substituting (\ref{3.16}) into (\ref{2.22}) and (\ref{2.23}), we have
\begin{equation}\label{3.17}
\mathcal{K}(\mathcal{F})=\frac{1}{a^2X}
\end{equation}
and
\begin{equation}\label{3.18}
\mathcal{H}(\mathcal{F})=\frac{1}{a^2X^2}
\end{equation}
where the parameter $a^2=(c_1/2c_0^3)^{\frac{1}{2}}$. Thus, we obtain the expression $\mathcal{K}^{-1}(y)$ by (\ref{2.20}),
\begin{equation}\label{3.19}
\mathcal{K}^{-1}(y)=\frac{1}{a^2y}.
\end{equation}
Similarly, we also obtain the expression $\mathcal{H}(\mathcal{K}^{-1}(y))$ by (\ref{3.18}) and (\ref{3.19})
\begin{equation}\label{3.20}
\mathcal{H}(\mathcal{K}^{-1}(y))=a^2y^2.
\end{equation}
Thereupon, (\ref{2.21}) can be reduced to
\begin{equation}\label{3.28}
\frac{r}{2}y'+y-a^2y^2=0,
\end{equation}
which is provided with the solution
\begin{equation}\label{3.29}
y=\frac{1}{a^2-u_0^2r^2}.
\end{equation}
Here, $u_0$ is an integral constant. We have the solution for this system as follows
\begin{eqnarray}\label{3.30}
\alpha&=&\beta^{-1}=1-\varepsilon^2-\frac{r_s}{r}-\frac{a\varepsilon^3}{\sqrt{2}mr}\ln{\frac{|u_0r-a|}{u_0r+a}},\nonumber\\
\phi^0&=&t\pm\int{\frac{dr}{\alpha}[1-\alpha(1-\frac{u_0^2}{a^2}r^2)]^{\frac{1}{2}}},\nonumber\\
\phi^i&=&(\frac{c_0}{2c_1})^{\frac{1}{4}}x^i,
\end{eqnarray}
where $\varepsilon^2=\frac{m^2}{2u_0^2}$ and $r_s$ is another integral constant. (\ref{3.30}) gets a new type of spherically symmetric metric which has interesting properties. Let us discussed its behavior and astrophysical consequences in Section 5.

\subsection{Classification of the functions $\mathcal{F}$}
The functions $\mathcal{F}\in\mathcal{R}^{\mathcal{F}}$ may be classified under three types according to metric form, that is to say, they are classified by the relation between $\mathcal{K}(\mathcal{F})$ and $\mathcal{H}(\mathcal{F})$. In the special case, the solution of furry black hole is reduced to Schwarzschild solution ($S=0$), Sch-AdS and Sch-dS solutions ($\lambda=-2, S\neq 0$). Three types are

Type (i)\ \ : In the case of $\mathcal{K}(\mathcal{F})=\mathcal{H}(\mathcal{F})$, there are only Schwarzschild, Sch-AdS \\ \indent\ \ \ \ \ \ \ \ \ \ \ \ \ \ \ and Sch-dS solutions.

Type (ii)\ : In the case of $\mathcal{K}(\mathcal{F})=c\mathcal{H}(\mathcal{F})$, there are the solutions with metric of\\ \indent\ \ \ \ \ \ \ \ \ \ \ \ \ \ \  furry black hole.

Type (iii): In the case of $\mathcal{K}(\mathcal{F})\neq c\mathcal{H}(\mathcal{F})$, there are the solutions beyond the metric\\ \indent\ \ \ \ \ \ \ \ \ \ \ \ \ \ \   of furry black hole.

\subsection{Topological stability of St\"{u}ckelberg field configuration}
There is no way to smoothly deform the hedgehog configuration into configuration $\phi^i=0$ everywhere. Thus, it is topological stable. The simplest example is the 't Hooft-Polyakov monopole which appears when $SO(3)$ is broken to $U(1)$. The three-component Higgs field $\phi^i$ for the monopole takes hedgehog configuration \cite{Hooft, Polyakov}. The generalization of 't Hooft-Polyakov monopole is $SO(10)$ GUT solution which appears when $SO(10)$ is broken to $SU(3)\times U(1)$ \cite{Zhang}. Mathematically, topological classes of solutions can be identified with classes of homotopic two-sphere $S^2$. This is expressed by the statement that $\Pi_2[SO(3)/U(1)]=\Pi_1[U(1)]/\Pi_1[SO(3)]=Z/Z_2$ for the 't Hooft-Polyakov monopole. The tensor monopole has also been discussed in a LB theory \cite{Li0}.

We have chosen the function $\mathcal{F}$ depending on $W^{ij}$ through three combination $W_n=\textmd{Tr}W^n$ $(n=1,2,3)$ which will be necessary for the background to be invariant under the $SO(3)$ symmetry in the $\phi^i$ internal space \cite{Bebronne1,Comelli5}. In the case of $\phi^i=0$, LB massive gravity will degenerate into GR, and Lorentz symmetry is recuperated. We note that the homomorphic relation is well known between $SL(2, C)$ and the Lorentz group. Therefore, the universal covering group is $SU(2)\times SU(2)$ for the Lorentz group. Since
\begin{equation}\label{3.31}
\Pi_2[\frac{SU(2)\times SU(2)}{SO(3)}]\cong\Pi_1[SO(3)]=Z_2,
\end{equation}
the St\"{u}ckelberg field configuration (\ref{2.1}) corresponding to a monopole exists, and there is no way to smoothly deform the monopole into a configuration with Lorentz symmetry.

We note that the Minkowski solution (\ref{1.2}) is consistent with the monopole configuration $\phi^i=bx^i$. Therefore, the topological stability guarantees the existence of spherically symmetric solutions including the trivial one. We have to analyse the stability of a non-trivial spherically symmetric solution in the perturbed state in the next section.

\section{Stability of furry black holes}
Isolated furry black holes are nonsingular outside the event horizon and fully characterized by the parameters $r_s$ and $S$. However, a real furry black hole must not be fully described by these basic parameters and it is invariably in the perturbed state. We would know something about the interaction of black hole with its astrophysical environment so that we have to analyse the stability of furry black holes in the perturbed state.

There are three stages when a black hole is perturbed. First stage is the rapid response at very early time, on which the initial conditions have a great effect. Second stage is quasinormal ringing phase, whose characteristic oscillation frequencies and damping times depend strongly on the quasinormal modes (QNMs). The QNMs are determined completely by the parameters of system, so they would carry significant information about the background curvature of the intervening spacetime. Finally, there is a tail stage, which decays approximately as a power in time owing to backscattering off the spacetime curvature \cite{Kokkotas}. QNMs have been extensively studied for various black holes by different methods \cite{Konoplya} including the string black holes \cite{Li3}, the brane-world models \cite{Berti} and the acoustic black holes \cite{Xi}. In order to comprehend which of these furry solutions are stable, and thus could exist in nature, we should consider their QNMs. Stability is assured when all QNMs are damped.

\subsection{Classification of furry black holes}

Note that the spacetime with metric (\ref{3.15}) is asymptotically flat which tends to vacuum at spatial infinity if only $\lambda>0$. Furthermore, we only consider the solutions that there exists the event horizon. If so, we can divide the metrics into following types,
\begin{eqnarray}\label{4.2}
&\mathrm{Type\quad IA}:&\quad \lambda>0, r_s>0\quad\textmd{and}\quad S\geq0;\\
&\mathrm{Type\quad IB}:&\quad \lambda>1, r_s>0, S<0\quad\textmd{and}\quad r_s\geq\frac{\lambda}{\lambda-1}(S-\lambda S)^{\frac{1}{\lambda}};\\
&\mathrm{Type\quad IC}:&\quad 0<\lambda<1, r_s<0, S>0\quad\textmd{and}\quad r_s\geq\frac{\lambda}{\lambda-1}(S-\lambda S)^{\frac{1}{\lambda}};\\
&\mathrm{Type\quad IIA}:&\quad \lambda>1, r_s<0\quad\textmd{and}\quad S>0;\\
&\mathrm{Type\quad IIB}:&\quad 0<\lambda<1, r_s>0\quad\textmd{and}\quad S<0.
\end{eqnarray}
By the analysis of Komar integral and the numerical results of QNMs, we argue that Type I is stable and Type II is unstable. Complementarily, there is an event horizon iff $r_s\geq r_{crit}$ for the Type IB and IC, and the critical value $r_{crit}=\frac{\lambda}{\lambda-1}(S-\lambda S)^{\frac{1}{\lambda}}$.

\subsection{Komar integral for furry black holes}

The total energy (or mass) is a tricky notion in the massive gravity or GR, because energy-momentum is a tensor than a vector, and energy-momentum tensor $T_{\mu\nu}$ only describes the properties of matter, not those of gravitational field. Fortunately, one can still define a conserved total energy by the Komar integral \cite{Komar} if stationary spacetime is equipped with a timelike Killing vector field $\xi^\mu$. The Komar integral associated with $\xi^\mu$ can be written as
\begin{equation}\label{4.7}
E=\frac{1}{4\pi G}\int_{\partial\Sigma}d^2x\sqrt{\gamma^{(\partial\Sigma)}}n_\mu\sigma_\nu\nabla^{\mu}\xi^{\nu},
\end{equation}
where $\Sigma$ is the spacelike hypersurface, $n^\mu$ is the unit normal vector associated with $\Sigma$, the boundary $\partial\Sigma$ is equipped with metric $\gamma_{ij}^{(\partial\Sigma)}$, and $\sigma^\mu$ is outward-pointing normal vector associated with $\partial\Sigma$. When the integral (\ref{4.7}) is convergent we can use it as the definition of total mass in all stationary asymptotically flat spacetime.

For the furry metric (\ref{3.15}), the boundary $\partial\Sigma$ can be taken two-sphere typically, so that
\begin{equation}\label{4.8}
\gamma_{ij}^{(\partial\Sigma)}dx^idx^j=r^2(d\theta^2+\sin^2\theta d\phi^2)
\end{equation}
and the normal vectors, normalized to $n_\mu n^\mu=1$ and $\sigma_\mu\sigma^\mu=-1$, have nonzero components $n_0=\alpha^{\frac{1}{2}}$, $\sigma_1=-\alpha^{-\frac{1}{2}}$ with other components vanishing. Thus, we have $n_\mu\sigma_\nu\nabla^{\mu}\xi^{\nu}=-\nabla^{0}\xi^1$. The Killing vector is $\xi^\mu=(1,0,0,0)$, so we have $\nabla^{0}\xi^1=g^{00}\Gamma_{00}^1\xi^0=-\frac{1}{2}\alpha'$. Putting them all together, the Komar integral of a furry black hole is
\begin{eqnarray}\label{4.9}
E=\frac{1}{8\pi G}\int d\theta d\phi r^2\sin\theta\alpha'=\frac{1}{2G}(r_s+\frac{\lambda S}{r^{\lambda-1}}).
\end{eqnarray}
We note that the Komar integral (\ref{4.9}) still depends on the radial coordinate $r$, which demonstrates a remarkable characteristic of the furry black hole. Obviously, the Komar integral is only convergent if $\lambda\geq1$. If so, we can use (\ref{4.9}) as the definition of total energy (mass) when $r$ tends to infinite. On the contrary, we can't define total energy (mass) if $0<\lambda<1$, even though the Komar energy defined well at the spacial finite.

Crossing $\tilde{r}=(-\frac{\lambda S}{r_s})^\frac{1}{\lambda-1}$, the Komar integral would change sign so that the attractive gravitational field converts into replusion or \textit{vice versa}. For Type I case, the Komar integral is always positive definite on the outside of horizon $r_h$. For Type II case, the Komar integral is positive definite in $r_h<r<\tilde{r}$ and negative definite in $r>\tilde{r}$. It is known rigorously that the Schwarzschild black hole is stable against external perturbation \cite{Chandrasekhar}. The results for QNMs support this conclusion, as shown by the negative imaginary parts of QNMs \cite{Iyer}.The Schwarzschild black hole is a special case of Type IA whose Komar integral is positive definite. The Komar integral is also positive definite outside horizon for Type IB and IC. However, the Komar integral is negative definite for Type IIA and Type IIB in the region $r>\tilde{r}$, which is an essential distinction between Type I and Type II. Thus, we speculate that Type I is stable and Type II is unstable which can be verified by the results of QNMs for various furry black holes.

\subsection{Quasinormal modes of furry black holes}

The QNMs for spin 0 and 1 may be derived assuming the coupling between the spin 0 and 1 fields to gravity is the same as those in GR. Since Lorentz invariance has been broken by mass term and therefore the calculating QNMs is not obvious in the case of gravitational perturbations. The perturbation equation for a furry black hole can be reduced to Schr\"{o}dinger-like form for stationary background
\begin{equation}\label{4.10}
{\frac{{d^{2}\psi (r_{ \ast} )}}{{dr_{ \ast}^2} }} + [\omega ^{2}-V(r)]\psi(r_{ \ast }) = 0,
\end{equation}
where the tortoise coordinate $r_{\ast}$ maps the semi-infinite region from the horizon to infinite into ($-\infty,\infty$) region, and $r_{\ast}$ is defined as
\begin{equation}\label{4.11}
dr_{\ast}=\beta dr.
\end{equation}
The effective potential \cite{medved1,medved2}is
\begin{equation}\label{4.12}
V(r)=\frac{\alpha}{r^2}[l(l+1)+\alpha'(1-s^2)r+(s-s^2)(1-\alpha)],
\end{equation}
where $s=0$ and $1$ corresponds to the scalar field and Maxwell field, respectively. Here, $l$ is the multipole quantum number (the angular harmonic index). The effective potential has a maximum at $r_*=(r_*)_0$.

The WKB approximative method was first applied by Schutz and Will \cite{Schutz} to the problem of scattering around black holes. This method is based on matching of the asymptotic WKB solutions at spatial infinity and the event horizon with the Taylor expansion near the top of the effective potential barrier through the two turning points. Here, we adopt the third-order WKB method \cite{Iyer}. The QNMs is as follows
\begin{equation}\label{4.13}
\omega^2=[V_{0}+(-2V_{0}^{''})^{\frac{1}{2}}\Lambda]-i\nu(-2V_{0}^{''})^\frac{1}{2}(1+\Omega),
\end{equation}
where the subscript $0$ on a variable denotes the value of the variable at $(r_{\ast})_0$ ($V_0''\neq0$), and $\nu\equiv n+\frac{1}{2}$ (overtone number $n=0,1,2,\cdots$, for $Re(\omega)>0$). Furthermore,
\begin{eqnarray}
\Lambda(n)&=&\frac{1}{(-2V^{''}_0)^{1/2}}\left\{\frac{1}{8}\left(\frac{V^{(4)}_0}{V^{''}_0}\right)
\left(\frac{1}{4}+\nu^2\right)-\frac{1}{288}\left(\frac{V^{'''}_0}{V^{''}_0}\right)^2
(7+60\nu^2)\right\}\ \label{4.14},\\
\Omega(n)&=&\frac{1}{(-2V^{''}_0)^{1/2}}\bigg\{\frac{5}{6912}
\left(\frac{V^{'''}_0}{V^{''}_0}\right)^4
(77+188\nu^2)-
\frac{1}{384}\left(\frac{V^{'''^2}_0V^{(4)}_0}{V^{''^3}_0}\right)
(51+100\nu^2)\nonumber\\
&+&\frac{1}{2304}\left(\frac{V^{(4)}_0}{V^{''}_0}\right)^2(67+68\nu^2)
+\frac{1}{288}
\left(\frac{V^{'''}_0V^{(5)}_0}{V^{''^2}_0}\right)(19+28\nu^2)\nonumber\\&-&\frac{1}{288}
\left(\frac{V^{(6)}_0}{V^{''}_0}\right)(5+4\nu^2)\bigg\},\label{4.15}
\end{eqnarray}
where the primes and the superscript ($j$) denote differentiation with respect to $r_{\ast}$. The numerical results of QNMs for electromagnetic perturbation ($s=1$) are listed in Tables 1-6. As a reminder, the results of QNMs verifies our conjecture: Type I is stable and Type II is unstable.

It is worthy that the corollary of QNMs are the same as those for the massless $s=2$ perturbation. However, we will run up against difficulties if we use WKB method to the scalar perturbation ($s=0$). In Type II and $s=0$ case, the effective potential (\ref{4.12}) is not a potential barrier for lower $l$ so that it can't reduce to the problem of scattering around black hole.

For Type I furry black holes, the real parts of the quasinormal frequencies increase as $l$, which means that the large $l$ is, the faster black hole oscillates; the imaginary parts are always negative, which corresponds to the oscillation of the black hole decays. Thus, Type I black hole will tend to stable eventually. For Type II furry black holes, there exist the positive imaginary parts for the lower modes so that they are unstable. In Tables 1-8, we have taken $|r_s|=1$, therefore, the unite of QNMs is $|r_s|^{-1}\approx1.02\times10^5(M_\odot/M)\textmd{sec}^{-1}$. For $(l,n)=(1,0)$ quasinormal mode of Type IIA furry black hole with $|M|=M_\odot$ in Table 5, the electromagnetic perturbation will enhance 42 times per microsecond. Obviously, the life of unstable black hole is proportional to its mass.
\begin{table}
\caption{QNMs of Type IA furry black hole for electromagnetic perturbation with $r_s>0$, $S=\frac{1}{2}r_s^2$ and $\lambda=2$. Here, we have worked in unite with $|r_s|=1$.}
\begin{indented}
\item[]\begin{tabular}{ccccc}
 \hline
 $l$ &$n=0$& $n=1$&$n=2$&$n=3$\\ \hline
1&0.3782-0.1643\textit{i}&0.3007-0.5316\textit{i}&0.1935-0.9198\textit{i}&0.0480-1.3174\textit{i}\\
2&0.7207-0.1685\textit{i}&0.6725-0.5190\textit{i}&0.5970-0.8888\textit{i}&0.5010-1.2685\textit{i}\\
3&1.0404-0.1698\textit{i}&0.0059-0.5161\textit{i}&0.9466-0.8758\textit{i}&0.8704-1.2463\textit{i}\\
4&1.3537-0.1703\textit{i}&1.3269-0.5150\textit{i}&1.2784-0.8690\textit{i}&1.2141-1.2327\textit{i}\\
5&1.6644-0.1706\textit{i}&1.6424-0.5145\textit{i}&1.6015-0.8651\textit{i}&1.5458-1.2238\textit{i}\\
\hline
\end{tabular}
\end{indented}
\end{table}
\begin{table}
\caption{QNMs of Type IA furry black hole for electromagnetic perturbation with $r_s>0$, $S=r_s^\frac{1}{2}$ and $\lambda=\frac{1}{2}$. Here, we have worked in unite with $|r_s|=1$.}
\begin{indented}
\item[]\begin{tabular}{ccccc}
 \hline
 $l$ &$n=0$& $n=1$&$n=2$&$n=3$\\ \hline
1&0.1656-0.0507\textit{i}&0.1512-0.1579\textit{i}&0.1310-0.2703\textit{i}&0.1050-0.3844\textit{i}\\
2&0.3000-0.0514\textit{i}&0.2913-0.1563\textit{i}&0.2769-0.2648\textit{i}&0.2590-0.3759\textit{i}\\
3&0.4287-0.0516\textit{i}&0.4225-0.1559\textit{i}&0.4113-0.2624\textit{i}&0.3968-0.3714\textit{i}\\
4&0.5557-0.0517\textit{i}&0.5508-0.1557\textit{i}&0.5418-0.2613\textit{i}&0.5296-0.3687\textit{i}\\
5&0.6819-0.0517\textit{i}&0.6780-0.1556\textit{i}&0.6704-0.2606\textit{i}&0.6599-0.3671\textit{i}\\
\hline
\end{tabular}
\end{indented}
\end{table}
\begin{table}
\caption{QNMs of Type IB furry black hole for electromagnetic perturbation with $r_s>0$, $S=-\frac{1}{5}r_s^2$ and $\lambda=2$. Here, we have worked in unite with $|r_s|=1$.}
\begin{indented}
\item[]\begin{tabular}{ccccc}
 \hline
 $l$ &$n=0$& $n=1$&$n=2$&$n=3$\\ \hline
1&0.6082-0.1867\textit{i}&0.5541-0.5836\textit{i}&0.4796-1.0012\textit{i}&0.3851-1.4244\textit{i}\\
2&1.1064-0.1908\textit{i}&1.0745-0.5798\textit{i}&1.0222-0.9825\textit{i}&0.9571-1.3945\textit{i}\\
3&1.5819-0.1918\textit{i}&1.5594-0.5792\textit{i}&1.5192-0.9750\textit{i}&1.4669-1.3793\textit{i}\\
4&2.0510-0.1923\textit{i}&2.0336-0.5790\textit{i}&2.0012-0.9713\textit{i}&1.9573-1.3704\textit{i}\\
5&2.5172-0.1925\textit{i}&2.5030-0.5789\textit{i}&2.4760-0.9693\textit{i}&2.4384-1.3648\textit{i}\\
\hline
\end{tabular}
\end{indented}
\end{table}
\begin{table}
\caption{QNMs of Type IC furry black hole for electromagnetic perturbation with $r_s<0$, $S=\frac{5}{2}(-r_s)^{\frac{1}{2}}$ and $\lambda=\frac{1}{2}$. Here, we have worked in unite with $|r_s|=1$.}
\begin{indented}
\item[]\begin{tabular}{ccccc}
 \hline
 $l$ &$n=0$& $n=1$&$n=2$&$n=3$\\ \hline
1&0.0876-0.0196\textit{i}&0.0843-0.0599\textit{i}&0.0791-0.1017\textit{i}&0.0938-0.1440\textit{i}\\
2&0.1552-0.0197\textit{i}&0.1533-0.0596\textit{i}&0.1499-0.1003\textit{i}&0.1457-0.1418\textit{i}\\
3&0.2207-0.0198\textit{i}&0.2193-0.0595\textit{i}&0.2167-0.0998\textit{i}&0.2133-0.1406\textit{i}\\
4&0.2207-0.0198\textit{i}&0.2844-0.0595\textit{i}&0.2824-0.0995\textit{i}&0.2795-0.1400\textit{i}\\
5&0.3501-0.0198\textit{i}&0.3492-0.0595\textit{i}&0.3475-0.0994\textit{i}&0.3451-0.1396\textit{i}\\
\hline
\end{tabular}\\
\end{indented}
\end{table}
\begin{table}
\caption{QNMs of Type IIA furry black hole for electromagnetic perturbation with $r_s<0$, $S=\frac{1}{10}r_s^2$ and $\lambda=2$. Here, we have worked in unite with $|r_s|=1$.}
\begin{indented}
\item[]\begin{tabular}{ccccc}
 \hline
 $l$ &$n=0$& $n=1$&$n=2$&$n=3$\\ \hline
1&13.243+36.813\textit{i}&41.621+98.291\textit{i}&90.522+172.11\textit{i}&156.37+257.15\textit{i}\\
2&9.4002-24.533\textit{i}&9.6272+85.761\textit{i}&42.807+150.31\textit{i}&88.781+221.11\textit{i}\\
3&31.462-22.446\textit{i}&13.037-80.206\textit{i}&12.991+141.17\textit{i}&48.945+205.92\textit{i}\\
4&48.402-22.892\textit{i}&32.260-77.273\textit{i}&10.168-136.13\textit{i}&19.633+197.64\textit{i}\\
5&63.455-23.280\textit{i}&49.602-75.662\textit{i}&30.041-132.91\textit{i}&4.3625-192.43\textit{i}\\
\hline
\end{tabular}
\end{indented}
\end{table}

\begin{table}
\caption{QNMs of Type IIB furry black hole for electromagnetic perturbation with $r_s>0$, $S=-10r_s^{\frac{1}{2}}$ and $\lambda=\frac{1}{2}$. Here, we have worked in unite with $|r_s|=1$.}
\begin{indented}
\item[]\begin{tabular}{ccccc}
 \hline
 $l$ &$n=0$& $n=1$&$n=2$&$n=3$\\ \hline
1&1199.1+1913.6\textit{i}&2779.6+4486.8\textit{i}&5128.1+7685.5\textit{i}&8234.5+11451\textit{i}\\
2&591.93+1512.2\textit{i}&1551.1+3811.0\textit{i}&3161.7+6517.5\textit{i}&5367.0+9582.8\textit{i}\\
3&143.45+1235.1\textit{i}&801.08+3511.2\textit{i}&2042.4+6018.4\textit{i}&3783.2+8769.7\textit{i}\\
4&327.47-1018.3\textit{i}&229.45+3341.3\textit{i}&1247.2+5747.2\textit{i}&2691.9+8322.6\textit{i}\\
5&809.10-930.26\textit{i}&252.12-3233.6\textit{i}&618.35+5580.2\textit{i}&1855.5+8046.3\textit{i}\\
\hline
\end{tabular}
\end{indented}
\end{table}

\begin{table}
\caption{QNMs of MBH with the metric (\ref{3.30}), $\varepsilon=\frac{1}{100}$ and $\frac{a^2}{m^2}=\frac{r_s^2}{2\varepsilon^2}$. Here, we have worked in unite with $|r_s|=1$.}
\begin{indented}
\item[]\begin{tabular}{ccccc}
 \hline
 $l$ &$n=0$& $n=1$&$n=2$&$n=3$\\ \hline
1&0.4917-0.1862\textit{i}&0.4226-0.5916\textit{i}&0.3286-1.0180\textit{i}&0.2038-1.4509\textit{i}\\
2&0.9141-0.1901\textit{i}&0.8715-0.5818\textit{i}&0.8045-0.9915\textit{i}&0.7209-1.4110\textit{i}\\
3&1.3133-0.1912\textit{i}&1.2828-0.5795\textit{i}&1.2300-0.9799\textit{i}&1.1627-1.3908\textit{i}\\
4&1.7058-0.1917\textit{i}&1.6820-0.5786\textit{i}&1.6389-0.9738\textit{i}&1.5817-1.3782\textit{i}\\
5&2.0954-0.1919\textit{i}&2.0760-0.5781\textit{i}&2.0396-0.9703\textit{i}&1.9900-1.3699\textit{i}\\
\hline
\end{tabular}
\end{indented}
\end{table}

\begin{table}
\caption{QNMs of MBH with the metric (\ref{3.30}), $\varepsilon=\frac{1}{100}$ and $\frac{a^2}{m^2}=\frac{r_s^2}{2}$. Here, we have worked in unite with $|r_s|=1$.}
\begin{indented}
\item[]\begin{tabular}{ccccc}
 \hline
 $l$ &$n=0$& $n=1$&$n=2$&$n=3$\\ \hline
1&0.4917-0.1862\textit{i}&0.4226-0.5916\textit{i}&0.3286-1.0180\textit{i}&0.2039-1.4509\textit{i}\\
2&0.9142-0.1901\textit{i}&0.8716-0.5818\textit{i}&0.8046-0.9915\textit{i}&0.7210-1.4110\textit{i}\\
3&1.3134-0.1912\textit{i}&1.2828-0.5795\textit{i}&1.2301-0.9799\textit{i}&1.1628-1.3908\textit{i}\\
4&1.7059-0.1917\textit{i}&1.6821-0.5786\textit{i}&1.6390-0.9738\textit{i}&1.5818-1.3782\textit{i}\\
5&2.0956-0.1919\textit{i}&2.0761-0.5781\textit{i}&2.0398-0.9703\textit{i}&1.9901-1.3700\textit{i}\\
\hline
\end{tabular}
\end{indented}
\end{table}

\section{Phenomenological consequences}
\subsection{A new astronomical object and its behavior}
The behavior of the metric (\ref{3.30}) is determined by the two integral constants $r_s$ and $u_0$, and the values of parameters $c_0$ and $c_1$. At the origin $r=0$ the term proportional to $r_s$ is singular, so the metric always possesses a singularity unless $r_s=0$. If we take $r_s=0$, the metric tends to $1$ since the logarithmic term becomes $m^2/2u_0^2$ at $r=0$. In the case of $r\gg1$, $\alpha$ tends to $1-(m^2/2u_0^2)$ which describes as a metric with a solid angular deficit \cite{Shi, Li1, Li2, Jin} since the logarithmic term is $r^{-2}$ order as $r\rightarrow\infty$.

There is another singularity at $r=a/u_0$ for the metric field $\alpha$. The point $r=a/u_0$ is a pole of order 2 for the scalar curvature $R$. These singularities may or may not be hidden by the horizon depending on the values of $r_s$ and $u_0$. The solutions possessing the horizon are candidates for modified black holes (MBH).

The horizon is always present if the integral constant $r_s\geq r_{crit}$ for fixed  $u_0$, $a$ and $m$ (see Figure 1), where
\begin{equation}
r_{crit}=\frac{a}{u_0}[1+\frac{\sqrt{2}\varepsilon^3u_0}{(1-\varepsilon^2)m}]^{\frac{1}{2}}.
\end{equation}
Actually, it is easy to prove the above-mentioned result by the asymptotical behavior of gravitational field.

Next, let us discuss stability of MBH with the metric (\ref{3.30}). Using (\ref{3.30}) and (\ref{4.13})-(\ref{4.15}), we list the results of QNMs in the Tables 7 and 8, which show this MBH is stable.
\begin{figure}
\begin{center}
\includegraphics[height=2.5in,width=3.1in]{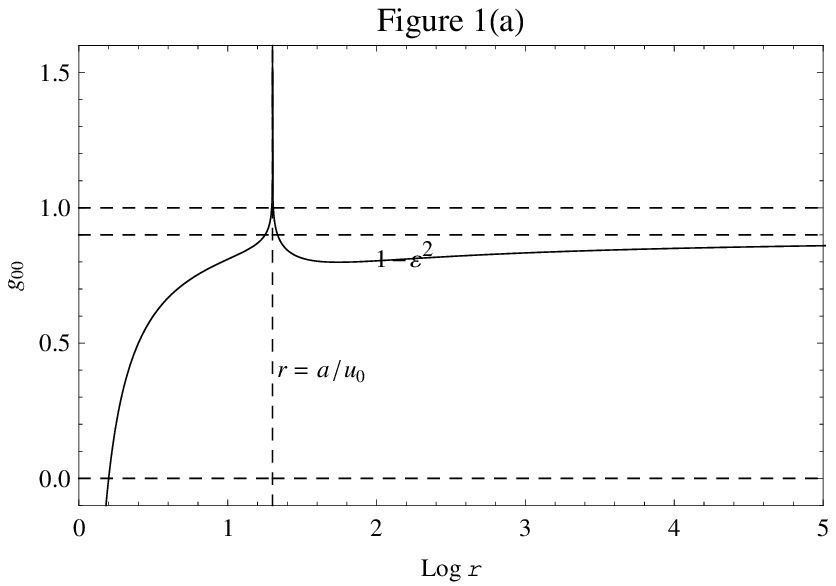}
\end{center}
\end{figure}
\begin{figure}
\begin{center}
\includegraphics[height=2.5in,width=3.1in]{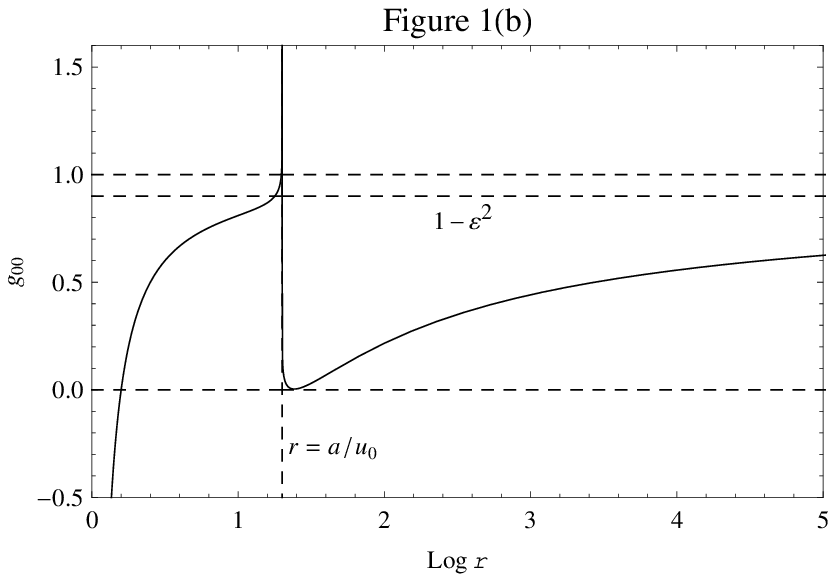}
\end{center}
\end{figure}
\begin{figure}
\begin{center}
\includegraphics[height=2.5in,width=3.1in]{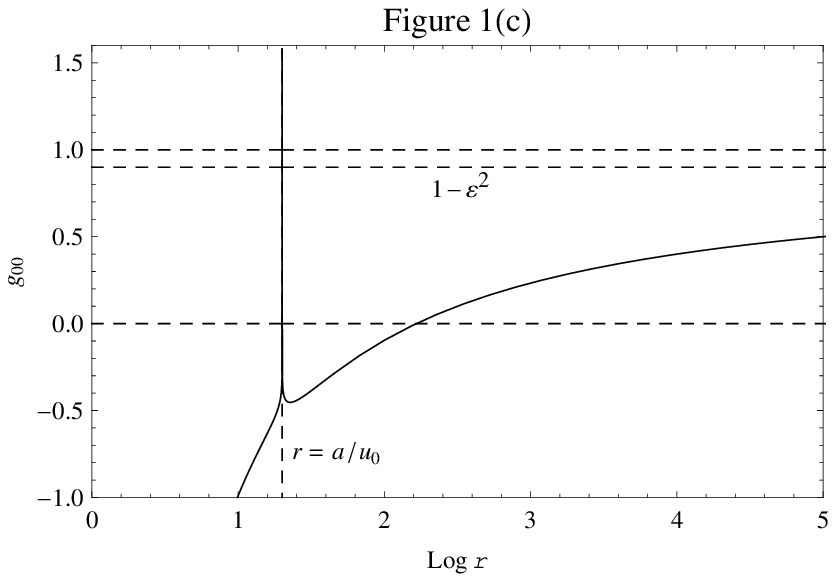}
\end{center}
\caption{\label{fig:i}The existence of horizon is examined for the metric (\ref{3.30}). Figure (a) corresponds to naked singularity at $r=a/u_0$ with the integral constant $r_s<r_{crit}$. The horizon $r_h$ is shown in the figures (b) and (c) where the integral constant $r_s=r_{crit}$ (b) and $r_s>r_{crit}$ (c).}
\end{figure}
\subsection{The deflection of a light ray and the perihelion shift of Mercury}
For a given gravitational field of the astronomical object with the metric (\ref{3.30}), the equation of motion then permits us to predict the trajectories of light signals, which we can compare with observation data. In the case of $\varepsilon\ll 1$, we should be able to neglect the $\varepsilon^3$ term in (\ref{3.30}) then the geodesic equations in the plane $\theta=\pi/2$ are
\begin{eqnarray}
\frac{dt}{dp}=\frac{1-\varepsilon^2}{\alpha(r)},\label{5.1}\\
r^2\frac{d\phi}{dp}=J,\label{5.2}\\
A(r)(\frac{dr}{dp})^2=\frac{1-\varepsilon^2}{\alpha(r)}-\frac{J^2}{r^2}-E^2,\label{5.3}
\end{eqnarray}
where $p$ is an affine parameter to coincide with the coordinate $t$ asymptotically, and the integral constant $E$ is such that $ds^2=-E^2dp^2$. $J$ is also a constant of motion, the angular momentum. By rescaling the following quantities,
\begin{eqnarray}
r_s=\tilde{r_s}(1-\varepsilon^2),\phi=\tilde{\phi}(1-\varepsilon^2)^{-\frac{1}{2}},J=\tilde{J}(1-\varepsilon^2)^{\frac{1}{2}},\nonumber\\
E^2=\tilde{E}^2(1-\varepsilon^2),t=\tilde{t}(1-\varepsilon^2)^{-1},p=\tilde{p}(1-\varepsilon^2)^{-1},\label{5.4}
\end{eqnarray}
(\ref{5.1})-(\ref{5.3}) look the same as those in the ordinary Schwarzschild metric. Therefore, we can use results of the Schwarzschild case for our issues. In a Schwarzschild metric, the change in the angular coordinate for a light ray scattered is $\Delta\tilde{\phi}=\pi+2r_sr_0^{ -1}$, where $r_0$ is the closest distance of the light ray from the centre of gravitational field. The angle by which light is deflected for this gravitational field is
\begin{equation}
\Delta\phi=\pi(1-\varepsilon^2)^{-1/2}+2r_s(1-\varepsilon^2)^{-3/2}r_0^{-1}-\pi.\label{5.5}
\end{equation}

One of the classical Einstein effects is the deflection of a light ray in the gravitational field of the sun. The effect is small but large enough to the detectable, which has been tested by observations during total eclipses of the sun on the apparent positions of stars whose light has passed close to the limb of the sum. The result of optical measurements is consistent with the prediction of GR, but hardly a precision conformation. However, the optical measurement can be substituted with the radio interferometry in these measurements, which has led to the development of very long baseline interferometry (VLBI). Using radio transmission from certain quasars and measuring the deflection as the source is eclipsed by the sun, the most precise result is that $\Delta\phi_{exp}/\Delta\phi_{GR}=1.0001\pm0.0001$ in terms of a ratio of the observed deflection and the theoretical value $\Delta\phi_{GR}=4GM_\odot/R_\odot\approx8.488\times10^{-6}$ rad for GR \cite{Robertson}. Consider the case of $r_0=R_\odot$ in (\ref{5.5}), we can write $\varepsilon^2$ in terms $\Delta\phi_{GR}$ and $\Delta\phi_{exp}$ as follows
\begin{equation}\label{5.6}
\varepsilon^2=1-\{[\frac{1}{2}(\frac{\pi}{\Delta\phi_{GR}}+\frac{\Delta\phi_{exp}}{\Delta\phi_{GR}})+\sqrt{A}]^{\frac{1}{3}}+[\frac{1}{2}(\frac{\pi}{\Delta\phi_{GR}}+\frac{\Delta\phi_{exp}}{\Delta\phi_{GR}})-\sqrt{A}]^{\frac{1}{3}}\}^{-2},
\end{equation}
where
\begin{equation}\label{5.7}
A=\frac{1}{4}(\frac{\pi}{\Delta\phi_{GR}}+\frac{\Delta\phi_{exp}}{\Delta\phi_{GR}})^2+\frac{1}{27}(\frac{\pi}{\Delta\phi_{GR}})^3.
\end{equation}
Using (\ref{5.6}) and $\Delta\phi_{exp}\leq1.0002\Delta\phi_{GR}$, we have the upper bound of the parameter $\varepsilon^2$, that is, $\varepsilon^2\leq1.081\times10^{-9}$.

There are other constraints from the solar system tests such as the perihelion shift of Mercury and the gravitational redshift. For the astronomical object with the metric (\ref{3.30}), the perihelion shift
\begin{equation}\label{86}
\Delta\phi=\frac{6\pi GM}{ac^2(1-e^2)(1-\varepsilon^2)^2},
\end{equation}
where $e$ is the eccentricity of the ellipse and $a$ is semi-major axis. On the other hand, the measured perihelion shift of Mercury has been known accurately. After the perturbing effects of the other planets have been accounted for, the excess $\Delta\phi_{GR}$ is known to about $10^{-3}$ from radar observations of Mercury between 1966 and 1990 \cite{will}. Therefore, we have the upper bound of the parameter $\varepsilon^2$, that is, $\varepsilon^2\leq1.498\times10^{-3}$. This constrain is weaker than one from the deflection of a light ray.

The geometric interpretation of $\varepsilon^2$ is a solid deficit angle of space. It is useful that the constraint is converted into a constraint on a more physically transparent parameter. We note that the dimension of $u_0$ is $[Length]$. In the $M_{pl}$ and $\Lambda$ terms, we have $u_0\geq25.84\Lambda^2 M_{pl}^{-1}$ from the solar system tests. The existence of black hole is the integral constant $r_s\geq r_{crit}$, which can be rewritten as
\begin{equation}\label{c1}
r_{crit}=(\frac{c_1}{2c_0^3})^{\frac{1}{4}}\frac{\varepsilon M_{pl}}{(1-\varepsilon^2)\Lambda^2},
\end{equation}
by using the parameters in the Lagrangian (\ref{1.1}) such as $\Lambda$, $c_0$ and $c_1$.

\subsection{Einstein ring}
If the source of light, the astronomical object with the metric (\ref{3.30}), and the observer are aligned exactly, all the rays that pass at the appropriate impact parameter around this object, at any azimuth, reach the position of the observer. Under these special circumstance, the observer sees an infinite number of images, which form a ring around this object. Assuming the source is much farther from this object than the observer, the rays incident on this object are then nearly parallel to the line of the alignment, and the deflection angle required for the ray to reach the observer is $b/D$, where $D$ is the distance from this object to the observer. Thus, the angular radius of the Einstein ring is
\begin{equation}
\frac{b}{D}=\frac{\varepsilon^2\pi}{2}+\frac{2r_s}{b},
\end{equation}
which can also be written as
\begin{equation}
\frac{b}{D}=\frac{\varepsilon^2\pi}{4}+\sqrt{\frac{\varepsilon^4\pi^2}{16}+\frac{2r_s}{D}}.
\end{equation}
If $\varepsilon^2=O(10^{-9})$, we obtain
\begin{equation}
\frac{b}{D}\geq0.1 \textmd{arcsec}.
\end{equation}
In the Schwarzschild case, the deflection angle tends to zero as $D$ tends to infinity. In our case, the deflection angle tends to $\varepsilon^2\pi/2$ as $D$ tends to infinity, that is to say, $b$ also tends to infinity with $D$ because the solid deficit angle extends to infinity. This angle is at the limit of resolution of optical telescopes, but it is well within the resolution attainable by radio telescopes. Such an Einstein ring should be observable with ratio telescopes if they exist.

\subsection{Cosmological density constraint}
In order to obtain a cosmological constraint on the abundance of this astronomical object, we use the upper bound of the density parameter $\Omega_{s,0}=\rho_s/\rho_{crit,0}$. Larson and Hiscock have estimated the astrophysics bounds on cosmic strings by the contribution of energy density \cite{Larson}. Similarly, we can estimate the minimum separation between these astronomical objects. If the typical distance between these objects is defined to $2L$, the average density up to the $O(\varepsilon^2)$ order is
\begin{equation}\label{92}
\rho_s\approx\frac{1}{2G}(\frac{r_s+\varepsilon^2(L-R_s)}{4\pi L^3/3}),
\end{equation}
where $R_s$ is the radius of this astronomical object. From $\rho_{s,0}<\rho_{crit,0}$ and $r_s$, $R_s\ll L$, we have
\begin{equation}\label{93}
L>\varepsilon\Omega_{m,0}^{-\frac{1}{2}}H_0^{-1}.
\end{equation}
where $H_0$ is Hubble parameter at present. Obviously, the separation for these objects is more larger than the size of the Galaxy, so it is not easy to find these objects if they exist unless $\varepsilon^2$ is tiny ($\varepsilon^2<10^{-12}$).

\section{Conclusion and discussion}
In this paper we have developed a detailed study of the spherically symmetric solutions in Lorentz breaking massive gravity. We have shown clearly that all solutions of $d\mathcal{F}=0$ form a commutative ring $R^{\mathcal{F}}$ if the St\"{u}ckelberg field $\phi^i$ is taken as a hedgehog configuration $\phi^i=\phi(r)x^i/r$. Under the hedgehog configuration and $\mathcal{F}$ is a polynomial, there is the only solution $\phi=br$ for Einstein equations. Moreover, we have obtained the expression of solution to the functional differential equation with spherically symmetry if $\mathcal{F}\in R^{\mathcal{F}}$. Using this universal formula, we give known solutions including the Schwarzschild, dS and AdS if we take $\mathcal{F}\in S^{\mathcal{F}}$ where $S^{\mathcal{F}}\subset R^{\mathcal{F}}$ and $\partial\mathcal{F}/\partial X=0$. When $\mathcal{F}\in R^{\mathcal{F}}$ but $\mathcal{F}\notin S^{\mathcal{F}}$, we have given some analytical examples including the furry black holes and a new metric solution (\ref{3.30}). If we take $r_s\geq r_{crit}$ in the metric (\ref{3.30}), this solution will describe a modified black hole or star. The stability of the St\"{u}ckelberg field $\phi^i$ is guaranteed by the topological one. The stability of these black holes under perturbations are discussed using the analysis of Komar integral and the results of QNMs. We also discussed some phenomenological consequences for these solutions.

We have discussed the minimum separation between the astronomical objects from the cosmological restriction in detail. It shows that if the typical distance between these objects is defined to be $2L$, $L>\varepsilon\Omega_{m,0}^{-\frac{1}{2}}H_0^{-1}$. Obviously, it is not easy to find these objects because the separation for these objects is more larger than the size of the Galaxy.

Furthermore, one can consider a region where matter is present in the form of a perfect fluid with a constant energy density $\rho$ and pressure $p$. In other words, the novel features can be ascribed to the St\"{u}ckelberg fluid turned on by matter inside the body. Thus, a self-gravitating body can be described by matching the exterior with the interior solution. In comparison to GR, the resulting equations are difficult to solve analytically even for a constant matter fluid because of the presence of the St\"{u}ckelberg fields. Therefore, one has to rely upon a nonstandard perturbation expansion \cite{Comelli5}. It is an interesting question whether there is a star solution with pure St\"{u}ckelberg field. We will study this question further and it is hopeful that many observations will be done to test LB massive gravity.

\ack{This work is supported by National Science Foundation of China grant. No. 11205102 and Innovation Program of Shanghai Municipal Education Commission (12YZ089).}\\

\appendix

\section{St\"{u}ckleberg fields and subring}
The functionals $\mathcal{K}(\mathcal{F})$ and $\mathcal{H}(\mathcal{F})$ are defined as
\begin{eqnarray}
\mathcal{K}(\mathcal{F})&=&\mathcal{F}-2X\mathcal{F}_X,\label{A1}\\
\mathcal{H}(\mathcal{F})&=&\mathcal{F}+\frac{2\phi^2}{r^2}\mathcal{F}_1-\frac{4\phi^4}{r^4}\mathcal{F}_2+\frac{6\phi^6}{r^6}\mathcal{F}_3,\label{A2}
\end{eqnarray}
then the Einstein equations can be rewritten as
\begin{eqnarray}
\frac{\alpha'}{r}+\frac{\alpha-1}{r^2}-\frac{m^2}{2}\mathcal{K}(\mathcal{F})=0,\label{A3}\\
\frac{\alpha''}{2}+\frac{\alpha'}{r}-\frac{m^2}{2}\mathcal{H}(\mathcal{F})=0,\label{A4}\\
X\mathcal{F}_X+\frac{\phi'^2}{X}\mathcal{F}_1-\frac{2\phi'^4}{X^2}\mathcal{F}_2+\frac{3\phi'^6}{X^3}\mathcal{F}_3=0.\label{A5}
\end{eqnarray}
In this appendix, we shall prove the following lemmata using (\ref{A3})-(\ref{A5}), which describe substructure of $\mathcal{R}^{\mathcal{F}}$ and the St\"{u}ckleberg fields.

\textit{Lemma} 1. Let $\mathcal{F}=\sum a_{l_0l_1l_2l_3}(\frac{1}{X})^{l_0}w_1^{l_1}w_2^{l_2}w_3^{l_3}$ is a polynomial ($l_0, l_1, l_2, l_3\geq0$) and $\mathcal{F}\in\mathcal{R}^{\mathcal{F}}$, then $\phi^i=bx^i$ is the only solution for the spatial St\"{u}ckleberg fields under the static spherically symmetric ans\"{a}tz (\ref{2.1}). The solution $\mathcal{F}$ can be classified by the parameter $s=l_1+2l_2+3l_3$.

\textit{Proof}. The solution $\mathcal{F}$ is  generally a function of $X$, $w_i$ and $\phi'^2$ for (\ref{A5}), that is to say, $\mathcal{F}=\mathcal{F}(X, w_1, w_2, w_3, \phi'^2)$. However, $\mathcal{F}=\mathcal{F}(X, w_1, w_2, w_3)$ is necessary for the $SO(3)$ symmetry in the $\phi^i$ internal space \cite{Bebronne1}. Thus, $\phi^i=bx^i$ is the only solution for the spatial St\"{u}ckleberg fields under ans\"{a}tz (\ref{2.1}). In this case, (\ref{A5}) is reduced to an algebraic equation as follows
\begin{equation}\label{A6}
\sum(-1)^{l_1+l_3}a_{l_0l_1l_2l_3}b^{2s}(\frac{1}{X})^{l_0+s}M(X)(1+2X)^{l_1}(1+2X^2)^{l_2}(1+2X^3)^{l_3}=0
\end{equation}
where
\begin{equation}\label{A7}
M(X)=l_0+\frac{l_1}{1+2X}+\frac{2l_2}{1+2X^2}+\frac{3l_3}{1+2X^3},
\end{equation}
and $s=l_1+2l_2+3l_3$. Therefore, the polynomial solution $\mathcal{F}$ can be classified by $b^{2s}$.$\Box$

It is easy to find the linear independent $\mathcal{F}$ for $s=0,1,2,3$ as follows
\begin{eqnarray}
s=0:\quad \mathcal{F}=C_0^1;\label{A8}\\
s=1:\quad \mathcal{F}=C_1^1(\frac{b^2}{X}+w_1)+b^2\times(s=0\quad\textmd{solution});\label{A9}\\
s=2:\quad \mathcal{F}=C_2^1(w_1^2-w_2+4b^2w_1)+b^2\times(s=1\quad\textmd{solution});\label{A10}\\
\quad\quad\quad\quad\mathcal{F}=C_2^2(\frac{b^4}{X^2}-w_2)+b^2\times(s=1\quad\textmd{solution});\label{A101}\\
s=3:\quad \mathcal{F}=C_3^1(w_1^3-3w_1w_2-6b^4w_1+2w_3)+b^2\times(s=2\quad\textmd{solution});\label{A102}\nonumber\\
\\
\quad\quad\quad\quad\mathcal{F}=C_3^2(\frac{b^6}{X^3}+w_3)+b^2\times(s=2\quad\textmd{solution});\label{A103}\\
\quad\quad\quad\quad\mathcal{F}=C_3^3(\frac{b^6}{X^3}+\frac{2b^6}{X^2}+\frac{2b^6}{X}+w_1w_2)+b^2\times(s=2\quad\textmd{solution});\label{A104}\nonumber\\
\\
\quad\quad\quad\quad\mathcal{F}=C_3^4(\frac{b^6}{X^3}+\frac{6b^6}{X^2}+\frac{12b^6}{X}+w_1^3)+b^2\times(s=2\quad\textmd{solution}).\label{A105}\nonumber\\
\end{eqnarray}
where $C_i^j$ are constants. Using addition and multiplication of ring, we can obtain solution $\mathcal{F}$ for $s\geq4$.

The argument is given in the Lemma 1 that $\mathcal{F}=\mathcal{F}(X, w_1, w_2, w_3)$ is necessary for $SO(3)$ symmetry in the $\phi^i$ internal space. In other words, $\mathcal{F}$ cannot explicitly depend on $\phi'$. Although the quantities $w_{1,2,3}$ themselves depend on $\phi'$ (see (\ref{2.11})), $w_{1,2,3}$ are reduced to the functions of $X^{-1}$ iff $\phi=br$.

A reasonable theory of gravity ought to contain the known spherically symmetric solutions in GR (Schwarzschild, Sch-AdS or Sch-dS solution). From (\ref{A3}) and (\ref{A4}), we have
\begin{equation}\label{A106}
\frac{\alpha''}{2}-\frac{\alpha-1}{r^2}=m^2(X\mathcal{F}_X+\frac{\phi^2}{r^2}\mathcal{F}_1-\frac{2\phi^4}{r^4}\mathcal{F}_2+\frac{3\phi^6}{r^6}\mathcal{F}_3).
\end{equation}
Using (\ref{A5}), (\ref{A106}) can be written in the form
\begin{equation}\label{A107}
\frac{\alpha''}{2}-\frac{\alpha-1}{r^2}=m^2[(\frac{\phi^2}{r^2}-\frac{\phi'^2}{X})\mathcal{F}_1-2(\frac{\phi^4}{r^4}-\frac{\phi'^4}{X^2})\mathcal{F}_2+3(\frac{\phi^6}{r^6}-\frac{\phi'^6}{X^3})\mathcal{F}_3].
\end{equation}
In the case of $\phi=br$ and $X=1$, (\ref{A107}) is reduced to
\begin{equation}\label{A13}
\frac{\alpha''}{2}-\frac{\alpha-1}{r^2}=0,
\end{equation}
and
\begin{equation}\label{A14}
\alpha=1-\frac{r_s}{r}+\Lambda r^2,
\end{equation}
where $r_s$ and $\Lambda$ are integral constants.

In the aforesaid considerations, we have used the requirement of vacuum solutions in GR. Conversely, we should consider what is restriction of $\mathcal{F}$ if there is only the vacuum solution. We find that the constraint is simple $\mathcal{K}(\mathcal{F})=\mathcal{H}(\mathcal{F})$, and $S^{\mathcal{F}}=\{\mathcal{F}\in\mathcal{R}^{\mathcal{F}}\mid\mathcal{K}(\mathcal{F})=\mathcal{H}(\mathcal{F})\}$ forms a subring, $S^{\mathcal{F}}\subset\mathcal{R}^{\mathcal{F}}$. Here, we have chosen $\mathcal{F}=\sum a_{l_0l_1l_2l_3}(\frac{1}{X})^{l_0}w_1^{l_1}w_2^{l_2}w_3^{l_3}$ ($l_0, l_1, l_2, l_3\geq0$).

\textit{Lemma} 2. If we take $\mathcal{F}\in S^{\mathcal{F}}$, the Einstein equations (\ref{A3})-(\ref{A5}) have only the Schwarzschild, Sch-AdS or Sch-dS solution. $S^{\mathcal{F}}$ has an equivalence definition $S^{\mathcal{F}}=\{\mathcal{F}\in R^{\mathcal{F}}|\mathcal{F}_X=0\}$.

\textit{Proof}. Combining (\ref{A3}) and (\ref{A4}) with $\mathcal{K}(\mathcal{F})=\mathcal{H}(\mathcal{F})$, we find that $\alpha$ satisfies (\ref{A13}). Therefore, the Einstein equations have only the Schwarzschild, Sch-AdS or Sch-dS solution for any $\mathcal{F}\in S^{\mathcal{F}}$. Thus, $X$ must be $1$ and $\mathcal{H}(\mathcal{F})=\mathcal{K}(\mathcal{F})$ is equivalent to $\mathcal{F}_X=0$ from the Einstein equations.$\Box$

\textit{Lemma} 3. Let $\mathcal{F}=\sum a_{l_0l_1l_2l_3}(\frac{1}{X})^{l_0}w_1^{l_1}w_2^{l_2}w_3^{l_3}$ is a polynomial ($l_0, l_1, l_2, l_3\geq0$) and $S^{\mathcal{F}}=\{\mathcal{F}\in\mathcal{R}^{\mathcal{F}}\mid\mathcal{K}(\mathcal{F})=\mathcal{H}(\mathcal{F})\}$, then $S^{\mathcal{F}}$ forms a subring of $\mathcal{R}^{\mathcal{F}}$. The set $S^{\mathcal{F}}_c=\{\mathcal{F}\in\mathcal{R}^{\mathcal{F}}\mid\mathcal{K}(\mathcal{F})=c\mathcal{H}(\mathcal{F}),c\neq1\quad\textrm{is a fixed constant}\}$ forms a subgroup of the abelian group $\mathcal{R}^{\mathcal{F}^+}$.

\textit{Proof}. Suppose that $f^1, f^2 \in S^{\mathcal{F}}$ and $\mathcal{K}(f^i)=\mathcal{H}(f^i)$, ($i=1, 2$). From Lemma 2 and (\ref{A5}), we have $\mathcal{K}(f^1+f^2)=\mathcal{H}(f^1+f^2)$ and $\mathcal{K}(f^1f^2)=\mathcal{H}(f^1f^2)$ so $S^{\mathcal{F}}\subset\mathcal{R}^{\mathcal{F}}$. For $f^1, f^2\in S_c^{\mathcal{F}}$, we only have $\mathcal{K}(f^1+f^2)=c\mathcal{H}(f^1+f^2)$ so $S_c^{\mathcal{F}}\subset\mathcal{R}^{\mathcal{F}^+}$.$\Box$

Note that we always obtain new solutions beyond the furry black holes using the addition and multiplication for $\mathcal{F}^1\in S_{c_1}^{\mathcal{F}}$ and $\mathcal{F}^2\in S_{c_2}^{\mathcal{F}}$.
\\
\\
\\

\end{document}